\documentclass[11pt]{article}

\usepackage[utf8]{inputenc}
\usepackage[colorlinks=true, allcolors=blue]{hyperref}
\usepackage[margin=1 in]{geometry}
\usepackage{graphicx}
\usepackage{amsmath, amssymb, mathtools, mathrsfs, dsfont, amsthm}
\usepackage{enumerate}
\usepackage{fancyhdr}
\usepackage{authblk}
\usepackage{url}
\usepackage{tensor}
\graphicspath{/fig}
\usepackage{subcaption}

\title{\textbf{Cosmologies inside hyperbolic black holes}}
\author{Simon F. Ross\footnote{s.f.ross@durham.ac.uk}}
\affil{\textit{Centre for Particle Theory, Department of Mathematical Sciences Durham University, South Road, Durham DH1 3LE, U.K.}}
\date{\today}

\begin{document}
\maketitle
\begin{abstract}
Models with closed FRW cosmologies on the worldvolume of a constant-tension brane inside a black hole provide an interesting setup for studying cosmology holographically. However, in more than two worldvolume dimensions, there are limitations on such models with flat spatial slices. I show that these limitations can be avoided by considering instead hyperbolic slices. This also naturally makes contact with previous work on Euclidean wormholes.   
\end{abstract}
%\newpage
%\tableofcontents 

\section{Introduction}

\begin{figure}[ht]
\centering
    \includegraphics[width=.3\linewidth]{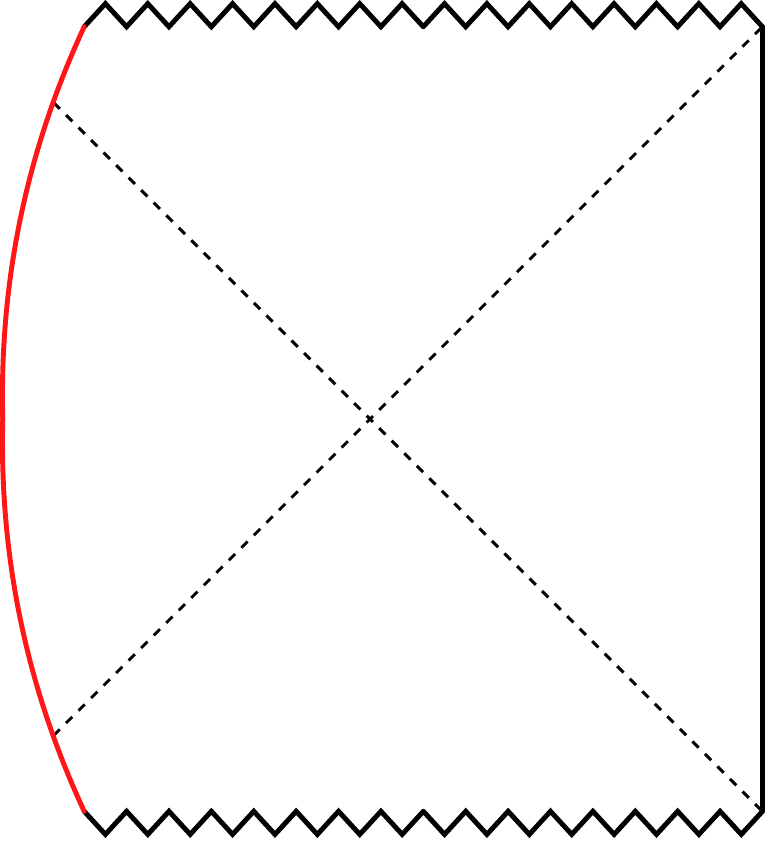}  
\caption{Penrose diagram of an AdS black hole with the left asymptotic region terminating in an ETW brane (shown in red). The worldvolume geometry of the ETW brane is a big-bang/big-crunch cosmology.}
\label{fig1}
\end{figure}

An approach to understanding closed universes with big-bang/big-crunch cosmologies holographically was proposed in \cite{Cooper:2018cmb} (and further developed in \cite{Antonini:2019qkt,VanRaamsdonk:2020tlr,Sully:2020pza,VanRaamsdonk:2021qgv,Antonini:2022blk,Antonini:2022xzo}). The idea is to consider an asymptotically AdS$_{d+1}$ black hole spacetime with a $d$-dimensional dynamical end of the world (ETW) brane behind the horizon providing an inner boundary of the spacetime, as depicted in figure \ref{fig1}. Starting from the $t=0$ surface, the ETW brane falls into the black hole and terminates at the singularity, so its worldvolume geometry is a big-bang/big-crunch cosmology. The state in the bulk on the $t=0$ surface is dual to some state in the dual $d$-dimensional CFT on the asymptotic boundary on the right in figure \ref{fig1}, which therefore includes a description of the cosmology on the ETW brane worldvolume. 

An appealing feature of this model is that the state in the bulk on the $t=0$ surface can be constructed from a Euclidean path integral. In the Euclidean section, the ETW brane moves outward away from $t=0$, and eventually meets the asymptotic boundary. Such solutions were proposed in \cite{Takayanagi:2011zk,Fujita:2011fp} as duals of boundary conformal field theories (BCFTs). The state in the $d$-dimensional CFT dual to the $t=0$ slice in the bulk is then constructed by starting with a $(d-1)$-dimensional boundary state specified by the BCFT and evolving through some period of Euclidean time. Such states have been extensively discussed in recent investigations of black holes, see e.g. \cite{Kourkoulou:2017zaj,Almheiri:2019hni,Penington:2019kki,Chen:2020tes}.

To have a controlled description of the cosmology on the ETW brane, we want to have a separation of scales between the scale that controls the curvature of the ETW brane and the bulk curvature, such that there is a good effective description of the dynamics in terms of ordinary Einstein gravity localised on the ETW brane. This can be achieved by taking the radial position $r_0$ of the brane at $t=0$ to be much larger than the horizon scale $r_H$, which can be achieved by increasing the tension $T$ of the ETW brane \cite{Randall:1999ee,Randall:1999vf,Karch:2000ct}. 

However, in simple examples of this construction this separation of scales is incompatible with the path integral construction of the state. If we are in $d>2$, and we take the bulk solution to be an uncharged black hole with flat spatial slices, when we increase $T$ for fixed bulk black hole geometry, there is a critical value $T=T_*$ at which the ETW brane intersects the asymptotic boundary at $t=0$. Increasing $T$ beyond $T_*$, the ETW brane will self-intersect before it reaches the asymptotic boundary. Thus, for $T> T_*$, we lose the Euclidean path integral construction of the state dual to the $t=0$ slice.\footnote{The $d=2$ case, where the bulk black hole is a BTZ solution \cite{Banados:1992wn}, is special; there the ETW brane always intersects the boundary at a quarter the period of the Euclidean solution.} One resolution of this problem, proposed in \cite{Antonini:2019qkt}, is to take the bulk solution to be a charged black hole. In \cite{VanRaamsdonk:2021qgv}, it was suggested that the problem could be avoided by introducing an additional interface brane, but in \cite{Fallows:2022ioc} (see also \cite{Waddell:2022fbn}) we found that the self-intersection problem persists in the presence of the interface brane. 

In this paper, we will see that another way to avoid this problem is to consider hyperbolic spatial slices, with constant negative curvature. Bulk black hole solutions with hyperbolic horizons were obtained in \cite{Birmingham:1998nr,Emparan:1998he,Emparan:1999gf}; these include a locally-AdS solution, which is a natural higher-dimensional analogue of the BTZ black hole. It is perhaps not surprising that considering ETW branes behind the horizon of these hyperbolic black hole then gives us a higher-dimensional scenario with behaviour similar to the $d=2$ case. 

Another way to see why considering hyperbolic spatial slices is helpful is to consider the worldvolume of the ETW brane. In the Euclidean black hole this is a wormhole; for hyperbolic spatial sections this is $\Sigma_g \times \mathbb R$, and we will see that when we consider the locally-AdS ``topological'' black hole, the worldvolume geometry is precisely the Maldacena-Maoz wormhole of \cite{Maldacena:2004rf}. This is a solution of Einstein's equations with a negative cosmological constant. By contrast, there is no classical solution describing a wormhole with flat spatial sections; the bulk spacetime thus plays a more essential role in the construction of the solutions where the ETW brane is a wormhole with flat spatial sections, and it seems natural that it's harder to construct solutions in the limit where the effective dynamics is approximated by Einstein gravity localised on the ETW brane.

As in the charged black hole case, these hyperbolic black holes have an extremal limit, where the temperature goes to zero for fixed horizon area. But the more important effect from the ETW brane perspective is that the curvature term dominates over the mass term in the metric at large distances. As a result, in the limit of interest, the mass term is negligible and the dynamics of the ETW brane is as in the locally-AdS case, which gives a finite time for the ETW brane to reach the Euclidean boundary, as in the $d=2$ BTZ black hole example. 

In the next section, we review some of the previous work on these models. In section \ref{hyp}, we introduce the generalization to consider hyperbolic spatial sections. Section \ref{disc} presents some concluding remarks. In appendix \ref{phase}, I include some speculations on the existence of other bulk saddle-points with the same boundary conditions as the hyperbolic black hole; this is a side comment independent of the main discussion. 

\section{End of the world brane cosmology}
\label{rev} 

The holographic model of cosmology we consider was first proposed in \cite{Cooper:2018cmb}. The model consists of an AdS black hole bulk with one asymptotic region, with a dynamical constant-tension ETW brane behind the horizon, as pictured in figure \ref{fig1}. The induced geometry on the ETW brane worldvolume is that of a closed FRW universe, with the radial position playing the role of the scale factor. The bulk action is 
\begin{equation}  
I = \frac{1}{16 \pi G}\left[ \int_\mathcal{M} d^{d+1} x \sqrt{-g}\, (R-2\Lambda) + 2\int_{\partial \mathcal{M}}  d^d y \sqrt{-h}\, K - 2(d-1) \int_{\mathcal{Q}} d^d y \sqrt{-h}\, T \right] ,
\end{equation}
where $\Lambda=-\frac{d(d-1)}{2 L^2}$ is a cosmological constant, $K$ is the trace of the extrinsic curvature and $T$ is the tension of the ETW brane with worldvolume $\mathcal{Q}$, which we take to be one component of the boundary $\partial\mathcal{M}$ of the spacetime, the other component corresponding to the asymptotically AdS conformal boundary. We consider a $(d+1)$-dimensional bulk spacetime, dual to a $d$-dimensional CFT on the boundary. 
Previous work considered an AdS-Schwarzchild black hole bulk solution
\begin{equation}
    ds^2 =  -f(r)dt^2+\frac{dr^2}{f(r)}+\frac{r^2}{L^2}dx^a dx_a, \quad \quad f(r)\equiv \frac{r^2}{L^2}-\frac{\mu}{r^{d-2}},
\end{equation}
where $a=1,\ldots,d-1$. This has a horizon at $r=r_H$, where $r_H^d = \mu L^2$. The brane has stress-energy tensor $8\pi G T_{ab}=(1-d)T h_{ab}$, and the action implies that the boundary condition for the bulk metric at $\mathcal{Q}$ is
\begin{equation}
    K_{ab}-Kh_{ab}=(1-d)T h_{ab}
\end{equation}
The $tt$ component of this equation leads to the brane equation of motion 
\begin{equation} \label{braneeom} 
   \left( \frac{d r}{d t}\right)^2=\frac{f^2(r)}{T^2 r^2}\left(T^2r^2 - f(r)\right).
\end{equation}
In the Lorentzian black hole geometry, the brane will reach a maximum radius $r_0$, with $(r_0)^d= \frac{r_H^d}{1 - T^2 L^2}$, which we take to occur at $t=0$. Note $r_0 > r_H$ for $T>0$, and $r_0 \to \infty$ as $T \to L^{-1}$. To the future and past of this, $r(t)$ decreases, as pictured in figure \ref{fig1}. The brane worldvolume geometry is thus a closed FRW big-bang/big-crunch cosmology, 
where the brane radius $r(t)$ plays the role of the scale factor, and the brane equation of motion \eqref{braneeom} corresponds to the Friedmann equation in this worldvolume cosmology. 

The state on the $t=0$ slice can be obtained by a Euclidean path integral. In the Euclidean black hole, the motion of the ETW brane is 
\begin{equation} \label{Euclidean braneeom} 
   \left( \frac{d r}{d \tau}\right)^2=\frac{f^2(r)}{T^2 r^2}\left(f(r) - T^2 r^2 \right).
\end{equation}
This now has a minimum at $r=r_0$. Since the ETW brane is inside the black hole in the Lorentzian geometry, it is at its minimum radius in the Euclidean solution at $\tau= \beta/2$, where $\beta = 2\pi L^2/r_H$ is the periodicity in Euclidean time $\tau$. It reaches the AdS boundary at a time  
\begin{equation} \label{tetw}
 \tau^{ETW} = \frac{\beta}{2} - \int_{r_0}^\infty \frac{dr}{f(r)} \frac{Tr}{\sqrt{f(r) - T^2 r^2}}.    
\end{equation}
To avoid self-intersections in the Euclidean solution, we need $\tau^{ETW}>0$. However, setting $r= r_0 x$, we have 
\begin{equation}
 \sigma^{ETW} = \frac{2 \tau^{ETW}}{\beta} = 1 - \frac{d}{2\pi} TL (y_0)^{\frac{d-2}{2}}  \int_1^\infty \frac{dx}{x^2 (1- (y_0)^{-d} x^{-d})  \sqrt{1-x^{-d}}},   
\end{equation}
where $y_0 = r_0/r_H$, so $(y_0)^{-d} = 1-T^2 L^2$. It is clear that for $d>2$ we can't take $r_0 \to \infty$ while keeping $\tau^{ETW} >0$. There must then be some critical value $T= T_* < L^{-1}$ such that $\tau^{ETW} = 0$, and if we consider $T > T_*$ we will have self-intersecting branes in the Euclidean solution.  

\section{Hyperbolic spatial sections}
\label{hyp} 

The previous discussion was for branes with flat spatial sections. Consider now the generalization to hyperbolic spatial sections. This changes relatively little in the previous analysis; the relevant black hole solution is now \cite{Birmingham:1998nr,Emparan:1998he,Emparan:1999gf} 
\begin{equation}
    ds^2 =  -f(r)dt^2+\frac{dr^2}{f(r)}+\frac{r^2}{L^2}ds^2_{\Sigma_g}, \quad \quad f(r) = \frac{r^2}{L^2} -1 -\frac{\mu}{r^{d-2}},
\end{equation}
where $\Sigma_g$ is a $(d-1)$-dimensional compact manifold of constant negative curvature. For $\mu \in (\mu_{min}, \infty)$, where 
\begin{equation}
  \mu_{min}  = - \frac{2}{d} \left( \frac{d-2}{d} \right)^{\frac{d-2}{2}} L^{d-2}, 
\end{equation}
the black hole has a horizon at $r=r_H>0$ where $f(r_H)=0$; we can write
\begin{equation}
  \mu = r_H^{d-2}  (\frac{r_H^2}{L^2} - 1 ).  
\end{equation}
The inverse temperature of the black hole is 
 \begin{equation} \label{bhyp} 
  \beta = \frac{4\pi r_H L^2}{d r_H^2 - (d-2) L^2} .    
\end{equation}
As $\mu \to \mu_{min}$, $r_H^2 \to \frac{d-2}{d} L^2$, and $f(r)$ develops a double root. The temperature goes to zero.\footnote{The possibility of having a vanishing temperature seems like it would be useful in constructing ETW brane solutions, but as we will see, the more important effect is the $-1$ in $f(r)$, which dominates over the mass term at large radial distances.}

For $\mu=0$, the solution is locally AdS$_{d+1}$. The metric with $\Sigma_g$ replaced by a non-compact hyperbolic space $H^{d-1}$ is simply AdS written in a Rindler-like coordinate system (the coordinate transformation between these coordinates and Poincar\'e coordinates basically reduces to the relation between Rindler coordinates and hyperbolic coordinates on the conformal boundary). The geometry of interest here is then obtained by quotienting by a discrete isometry group $\Gamma$ such that $H^{d-1}/\Gamma = \Sigma_g$. This $\mu=0$ solution is therefore referred to as the topological black hole. The $\mu=0$ solution has a finite temperature, which is related to the Rindler acceleration temperature in the boundary theory if we don't do the quotient. 

We consider an ETW brane inside this black hole, with a turnaround at $r=r_0$ at $t=0$. Considering the Euclidean continuation, the brane equation of motion is \eqref{Euclidean braneeom}, and the brane reaches the AdS boundary at 
\begin{equation} 
 \tau^{ETW} = \frac{\beta}{2} - \int_{r_0}^\infty \frac{dr}{f(r)} \frac{Tr}{\sqrt{f(r) - T^2 r^2}}.    
\end{equation}

The key difference from the previous flat case is the behaviour of the square root factor in the denominator: if we set $r= r_0 x$, in both cases, we pull out a factor of  $1/\sqrt{r_0^2 (1-T^2L^2)}$,  but whereas in the flat case we had $1-T^2 L^2 \sim 1/r_0^d$, so this factor goes like $r_0^{\frac{d-2}{2}}$, which blows up as $r_0 \to \infty$, for the hyperbolic black hole, $r_0^2 (1-T^2 L^2) = L^2 - \mu \frac{L^2}{r_0^{d-2}}$. The last term becomes negligible if $\mu$ remains finite as $r_0 \to \infty$, so $1/\sqrt{r_0^2 (1-T^2L^2)}$ has a finite limit as $r_0 \to \infty$. 

This can be simply illustrated by considering the case $\mu =0$. As noted above, we still have a finite temperature at $\mu=0$: \eqref{bhyp} gives $\beta = 2\pi L$. Setting $r=r_0 x$, we have
\begin{equation} 
 \tau^{ETW} = \pi L  - TL^2 \int_{1}^\infty \frac{x dx}{\left( x^2 - \frac{L^2}{r_0^2} \right)\sqrt{x^2-1}}. 
\end{equation}
As $TL \to 1$, $r_0 \to \infty$, and we can do the integral exactly; it's equal to $\frac{\pi}{2}$, so $\tau^{ETW} \to \frac{\pi L}{2}$. Thus, there's no self-intersection problem here; we can freely take the ETW brane to large radius in arbitrary dimensions.\footnote{The integrand is larger, and hence $\tau^{ETW}$ smaller, for finite $r_0$; but we are mostly interested in the region of large $r_0$, so we will not analyse in detail the positivity of  $\tau^{ETW}$ for finite $r_0$.}  It is interesting to note that for $\mu =0$, the bulk geometry is locally AdS$_{d+1}$, and the brane is locally AdS$_d$, so the solution we are considering here is just a quotient of the sub-critical braneworld introduced in \cite{Karch:2000ct}.

Generalizing to $\mu \neq 0$, so long as we hold $\mu$ fixed as we take the limit as $r_0 \to \infty$ the $\frac{\mu}{r^{d-2}}$ term in $f(r)$ becomes negligible in the limit, so the integral will converge to the above $\mu=0$ expression. The only difference for $\mu \neq 0$ is then that the black hole has a different temperature, so 
\begin{equation} 
 \tau^{ETW} = \frac{\beta}{2} - \frac{\pi L}{2}.   
\end{equation}
There is thus a good solution in the $r_0 \to \infty$ limit for all $\mu$ such that $\beta > \pi L$. This corresponds to $\mu < \mu_{max}$ where 
\begin{equation} 
\mu_{max} = (r_H^{max})^{d-2} \left( \frac{(r_H^{max})^2}{L^2} - 1 \right), \quad \frac{r_H^{max}}{L} =  \frac{2 + \sqrt{4 + d(d-2)}}{d}. 
\end{equation}
For fixed $T, L$, the boundary interval $\tau^{ETW}$ is determined by $\mu$; from the boundary perspective we would like to invert this relationship and fix $\tau^{ETW}$ and determine $\mu$. In the limit as $r_0 \to \infty$, it is easy to calculate that $\tau^{ETW}$ runs from $+\infty$ to $0$ monotonically as $\mu \in (\mu_{min}, \mu_{max})$. Thus, there is a one to one map from possible values of $\tau^{ETW}$ to $\mu$ in this range for large $r_0$. 

Consider the induced geometry on the ETW brane: in general this is 
\begin{equation} 
ds_b^2 = \frac{dr^2}{f(r)- T^2 r^2} + r^2 d\Sigma_g^2. 
\end{equation}
If $\mu=0$, setting $r=r_0 \cosh \rho$, we have
\begin{equation} 
ds_b^2 = r_0^2 (d \rho^2 + \cosh^2 \rho d\Sigma_g^2). 
\end{equation}
Thus, for the topological black hole, the induced geometry on the brane is precisely the Maldacena-Maoz wormhole \cite{Maldacena:2004rf}.   In general, if we take $r_0 \to \infty$ at fixed $\mu$, the mass term in $f(r)$ is negligible at the brane position, and the induced geometry on the brane will be approximately the same. This is natural, as the effective theory on the brane in this limit reduces to Einstein gravity, and the Maldacena-Maoz wormhole is a classical solution of this theory. This offers another perspective on the advantages of considering the brane with hyperbolic cross-sections.  

At finite $r_0$, the induced geometry on the brane is modified for $\mu \neq 0$; these modifications can be interpreted from the brane effective theory prespective as due to the effective stress tensor on the brane dual to the bulk geometry. Note that for $\mu <0$, this effective stress tensor will have a negative energy density. This can be interpreted as a Casimir energy due to putting the theory on a compact hyperbolic space. For $\mu=0$, the energy density on the brane vanishes due to a cancellation between the negative Casimir energy and the positive thermal energy associated with the finite temperature. 

\section{Discussion} 
\label{disc}

We have seen that considering spatial manifolds with negative curvature provides a good environment for implementing the holographic cosmology proposal of \cite{Cooper:2018cmb} in higher dimensions. The bulk solution is a hyperbolic black hole, whose structure in higher dimensions is more similar to the structure of the BTZ black hole in three dimensions. There are many questions about the holographic cosmology which it will be interesting to investigate in this context in the future: how the degrees of freedom on the brane are encoded in the boundary theory, what predictions these models make for the cosmological evolution on the brane, the role of the negative Casimir energy of the CFT, and others. 

In top-down models, the CFT usually includes scalar fields; conformally invariant scalar fields have a coupling to the curvature of the background which leads to an instability when the background is negatively curved. In the bulk this is related to an instability for probe branes at large radial position to run away towards the boundary \cite{Seiberg:1999xz}. This would seem to be a significant issue with the models we are considering, and indeed if we consider the zero temperature black hole the bulk solution is unstable; branes at any radial position run away to the boundary. However, if we consider the finite temperature solutions (for example in AdS$_5$) there is a non-trivial radial potential for probe (D3)-branes, and branes close to the horizon will collapse into the black hole \cite{Landsteiner:1999up}. Thus, branes must tunnel through this potential barrier to escape to infinity, and the dual CFT at finite temperature is meta-stable. The model should provide a useful way of describing the cosmology holographically on timescales short compared to the decay time.  

For flat spatial slices, the Euclidean ETW brane geometry can be analytically continued in one of the spatial directions to obtain an eternal traversable wormhole geometry. The relation between the problem of constructing ETW brane cosmologies with flat slices and the challenges in constructing eternal traversable wormholes was explored in \cite{VanRaamsdonk:2021qgv,Fallows:2022ioc,Waddell:2022fbn}, following \cite{Freivogel:2019lej}.\footnote{The construction of wormholes using a coupling between theories was earlier explored in \cite{Betzios:2019rds}. See also \cite{Betzios:2021fnm,Antonini:2022xzo,Antonini:2022opp} for more recent work on these wormholes.} It is therefore interesting to ask if there is a connection between the geometries with hyperbolic spatial slices studied here and Lorentzian wormholes. The metric on $\Sigma_g$ will not have a translation symmetry, but if it  has a surface of time reflection symmetry, we can define a Lorentzian solution by analytic continuation. This Lorentzian geometry will be a quotient of AdS similar to those considered in \cite{Horowitz:1998xk}, which gives again a time-dependent cosmology with big bang big crunch ``singularities" where the quotient degenerates. The spatial slices of this brane geometry will be a wormhole; whether it is traversable or not depends on the ratio of the time needed to cross the radial spatial direction between the two boundaries and the finite lifetime of the cosmology. In the simple $\mu=0$ case, both times should be of order one in units of the AdS scale, so understanding whether these wormholes are traversable seems to require a detailed case by case analysis. 

Finally, as we have emphasized connections to the $d=2$ case in this work, it is worth noting that for $d=2$ the Euclidean boundary conditions with an interval of Euclidean time between two brane boundaries can be filled in in two different ways: we can have a piece of the bulk BTZ black hole bounded by a connected ETW brane, or a piece of vacuum AdS bounded by disconnected ETW branes \cite{Fujita:2011fp}. We have discussed the analogue of the connected solution; is there an analogue of the disconnected one? This requires an analogue of thermal AdS for the hyperbolic black hole. No such solution is known, but in the appendix I suggest such solutions may exist. They are more difficult to construct analytically; it might be interesting to study them numerically. 

\section*{Acknowledgements}

I thank Mark van Raamsdonk for helpful discussions. This work is supported in part by STFC through grant ST/T000708/1, and by a grant from the Simons Foundation. This work was performed in part at the Aspen Center for Physics, which is supported by National Science Foundation grant PHY-1607611.

\appendix

\section{Phase transitions for hyperbolic black holes} 
\label{phase}

In the main body of the paper, we have considered adding ETW branes to the hyperbolic black holes introduced in \cite{Birmingham:1998nr,Emparan:1998he,Emparan:1999gf}. It is interesting to consider whether there might be other bulk solutions for the same boundary conditions, which could lead to a non-trivial phase structure analogous to the Hawking-Page phase transition \cite{Hawking:1982dh}.

The thermal partition function for a field theory on $S^d$ is given by a path integral on $S^1_\beta \times S^d$, where the size of the circle is fixed by the inverse temperature. There are two possible bulk saddle-points in the holographic description of this path integral: Euclidean Schwarzschild-AdS, and thermal AdS. In the first the $S^1$ is contractible in the interior of the spacetime, and in the second the $S^d$ is. There is an exchange of dominance between these two saddle-points which occurs at an order one value of the ratio of the size of the $S^1$ to the size of the $S^d$ \cite{Hawking:1982dh}. This corresponds to a phase transition in the thermal partition function, the Hawking-Page phase transition, which is interpreted in the dual as a deconfinement transition \cite{Witten:1998zw}.  

If we consider the field theory on $\mathbb R^d$, there is no phase transition; the dominant bulk saddle is the planar AdS black hole for all temperatures. However, if we consider compactifying the space to $T^d$, the story is more subtle. In supersymmetric theories, with supersymmetry-preserving boundary conditions on the torus, there is no phase transition, but if we consider antiperiodic boundary conditions on one or more cycles on the torus, there is a non-trivial phase structure. With antiperiodic boundary conditions on a spatial cycle, at zero temperature there is an AdS soliton solution \cite{Horowitz:1998ha}, which reproduces the expected negative Casimir energy of the field theory. At finite temperature we consider the field theory path integral on $S^1_\beta \times T^d$, and there are multiple bulk saddle-points: the planar black hole and thermal AdS solitons for any circles with antiperiodic boundary conditions. There is a phase transition when the ratio of the size of the thermal circle $S^1_\beta$ to the size of the smallest circle which is contractible in an AdS soliton is equal to one.\footnote{There are also transitions between different AdS solitons if we vary the relative size of different circles in the torus.}  

For the field theory on a compact hyperbolic space $\Sigma_g$, the finite temperature partition function is given by a path integral on $S^1_\beta \times \Sigma_g$. The only known bulk saddle is the hyperbolic black hole, and in the previous literature this has been taken as indicating the absence of a phase transition in this case. However, this may simply reflect our ignorance. By analogy to the torus case discussed above, we might expect that (assuming we consider antiperiodic boundary conditions for the fermions)  there would be bulk solutions where a non-contractible cycle in $\Sigma_g$ becomes contractible in the bulk. Indeed, if we consider a two-dimensional field theory on a Riemann surface $\Sigma_g$, such handlebody solutions have been explicitly constructed \cite{Faulkner:2013yia}. Constructing such solutions in the higher-dimensional case with  $S^1_\beta \times \Sigma_g$ boundary where $\Sigma_g$ has two or more dimensions will be challenging, as the solutions will only have a $U(1)$ symmetry, corresponding to translations along the $S^1$. The explicit constructions of \cite{Faulkner:2013yia} were only possible because the bulk solution was locally AdS$_3$, and in higher dimensions bulk vacuum solutions are not required to be locally AdS. But there is no obvious reason to expect such solutions not to exist. Indeed, given the experience with the torus case and Riemann surface boundaries in AdS$_3$, it would seem more surprising if they did not than if they did. 

If such solutions where a cycle in $\Sigma_g$ is contractible exist, there would then be multiple bulk saddle-points for the path integral on $S^1_\beta \times \Sigma_g$, and there could be phase transitions due to the competition between them and the hyperbolic black hole, as a function of the temperature and the moduli of $\Sigma_g$. There is however an interesting difference between this case and the torus: for the torus, the lowest-energy black hole has zero energy, while we expect the field theory to have a negative energy density (for antiperiodic fermions) at sufficiently low temperature due to the Casimir energy, so there is a clear argument that a different phase (the AdS soliton) must take over at low temperatures. For the theory on $\Sigma_g$, the low temperature black hole has negative energy, so this could already reproduce the Casimir energy in the field theory. Thus, even assuming that such solutions where a cycle in $\Sigma_g$ is contractible exist, the question of whether they dominate in the thermal partition function at low temperatures would remain open. This seems a very interesting question for future numerical investigation. 

Returning to the solutions with ETW branes considered in the body of the paper, if such solutions where a cycle in $\Sigma_g$ is contractible exist, for a field theory on an interval cross $\Sigma_g$ ended by brane boundaries there might then be solutions with disconnected ETW branes bounding a piece of such a solution.

\bibliographystyle{utphys}
\bibliography{bib}

\end{document}